# Modelling multiplex testing for outbreak control

Martyn Fyles[1,*], Christopher E. Overton[1,2], Tom Ward[1], Emma Bennett[1], Tom Fowler[3,4], Ian Hall[1,5]

1. UK Health Security Agency, Data, Analytics and Surveillance, 10 South Colonnade, London, UK
2. University of Liverpool, Department of Mathematical Sciences, Peach Street, Liverpool, UK
3. UK Health Security Agency, Clinical and Public Health, 10 South Colonnade, London, UK
4. Queen Mary University of London, William Harvey Research Institute, London, UK.
5. Department of Mathematics, University of Manchester, Manchester, UK

[*] Corresponding Author, Martyn.Fyles@ukhsa.gov.uk

## Abstract

During the SARS-CoV-2 pandemic, polymerase chain reaction (PCR) and lateral flow device (LFD) tests were frequently deployed to detect the presence of SARS-CoV-2. Many of these tests were singleplex, and only tested for the presence of a single pathogen. Multiplex tests can test for the presence of several pathogens using only a single swab, which can allow for: surveillance of more pathogens, targeting of antiviral interventions, a reduced burden of testing, and lower costs. Test sensitivity however, particularly in LFD tests, is highly conditional on the viral concentration dynamics of individuals. To inform the use of multiplex testing in outbreak detection it is therefore necessary to investigate the interactions between outbreak detection strategies and the differing viral concentration trajectories of key pathogens. Viral concentration trajectories are estimated for SARS-CoV-2, and Influenza A/B. Testing strategies for the first five symptomatic cases in an outbreak are then simulated and used to evaluate key performance indicators. Strategies that use a combination of multiplex LFD and PCR tests achieve; high levels of detection, detect outbreaks rapidly, and have the lowest burden of testing across multiple pathogens. Influenza B was estimated to have lower rates of detection due to its modelled viral concentration dynamics.

## Introduction

Outbreak detection for respiratory infections such as SARS-CoV-2 and influenza A/B is a common practice in settings that contain individuals that are either clinically vulnerable, or at higher risk of severe outcomes [1], such as care home residents. By detecting these outbreaks, interventions can be deployed to either prevent onwards transmission [1], or to deploy therapeutics such as antiviral medications [2]. The time it takes to detect an outbreak can be significantly impacted by the testing strategies and technologies used, which in turn impacts the efficacy of deployed interventions. In this paper, we explore different testing policies for the first five symptomatic infections in an outbreak setting for three different pathogens: SARS-CoV-2, influenza A and influenza B. This is achieved by developing models of test sensitivity over the course of an infection and then simulating outbreaks to evaluate key performance metrics. This can then be used to inform choices regarding the implementation of multiplex testing strategies in outbreak settings for SARS-CoV-2 and influenza A/B.

Multiplex tests analyse samples for the potential presence of multiple different pathogens using only a single test and swab. This contrasts with "singleplex" tests that can only test for the presence of a single pathogen per test and swab. Singleplex SARS-CoV-2 tests saw widespread use during the SARS-CoV-2 pandemic [3] when there was a single high-priority pathogen, however, as the SARS-CoV-2 crisis has abated and other key public health pressures such as influenza have returned to typical levels [4], singleplex testing policies may no longer be optimal in terms of efficacy, cost-effectiveness and burden of sample collection when several tests are required.

Currently, there exist two main multiplex testing technologies: real-time reverse transcription polymerase chain reaction (RT-PCR), and lateral flow devices (LFD). PCR tests are the gold standard in terms of test sensitivity and specificity; however, the swabs are typically processed in a laboratory which adds significant costs and delays due to the logistical requirements of transporting of the swab to a laboratory. Consequently, PCR testing in the community can have a turnaround time of several days, which is detrimental to the efficacy of outbreak control where speed of detection is a priority. LFD tests are self-contained test cassettes that do not require transportation to a laboratory and provide results in under 30 minutes. The ability of LFD tests to provide rapid results comes at the cost of a reduced sensitivity at lower viral concentrations when compared to a PCR test. This creates a complex testing landscape, where both technologies have relative strengths and weaknesses that need to be considered.

In line with guidance, it is common practice to test the first five symptomatic cases that occur, unless assessment of the outbreak suggests a different sampling approach is needed [1]. Once an outbreak is detected via a case testing positive interventions, that apply to the entire outbreak, are initiated that both prevent onward transmission and may administer therapeutics that improve clinical outcomes if administered early in the infection. This suggests that analysis of testing technologies and strategies should take an outbreak-level view, rather than an individual-level view. If individual-level sensitivity is used as the basis for analysis, the PCR test will always offer a significant advantage over LFD tests due to its high sensitivity. However, in an outbreak setting where multiple tests occur and only a single positive test result, along with an epidemiological link to other symptomatic cases, is needed to trigger outbreak-level interventions, it is likely that LFD testing will offer a comparable outbreak detection rate to PCR tests.

It has been demonstrated that test sensitivity is highly dependent on the viral concentration in swabs taken from individuals [5]. This is a proxy for the amount of virus present in a person's nasal or oral cavity. Consequently, it is necessary to understand the viral concentration of SARS-CoV-2 and influenza A/B cases over time, as this determines the test sensitivity profile of infection in an outbreak. To achieve this, viral concentration trajectories were modelled for SARS-CoV-2 and influenza A/B using publicly available datasets [6, 7]. This allows us to implement a model of LFD test sensitivity that depends on the viral concentration of the case at time of testing, based upon a model of SARS-CoV-2 test sensitivity.

Outbreaks are then simulated according to assumed values of the effective reproductive value and generation time distributions. This allows testing policies that use various combinations of LFD and PCR tests to be simulated, and for the key performance indicators of each strategy to be evaluated, such as the fraction of outbreaks detected, and the number of tests used in detecting the outbreak. Five different testing strategies for the first five symptomatic cases in an outbreak were implemented: an all-PCR testing strategy; an all-LFD testing strategy; a strategy where cases are first tested with an LFD test, and if negative retested with a PCR test; an all-LFD testing strategy with confirmatory PCR testing; and an all-LFD testing strategy with PCR retesting if the first five LFD results are all negative. The summaries of the key performance indicator statistics are used to draw conclusions regarding the expected performance of various multiplex testing strategies.

## Methods

### Viral concentration trajectory data and test sensitivity models

Two high quality datasets were used to estimate the viral concentration trajectories of SARS-CoV-2 and influenza A/B. For SARS-CoV-2, data were sourced from the Human Challenge Study [6]. The human challenge study performed daily longitudinal sampling of the viral concentration trajectory for 18 cases after deliberate infection with wild-type SARS-CoV-2. Swabbing was performed at two sites: the throat, and the mid-turbinate. For influenza A/B, the viral concentration trajectory data was obtained from Lau et al. [7], who performed daily longitudinal sampling of the viral concentration trajectory of the household members of index cases in Hong Kong 2009. This resulted in 26 cases of influenza A, and 18 influenza B cases. Both studies recorded the viral concentration in terms of the gene copies ml$^{-1}$.

A scoping literature search was conducted to identify other viable datasets with suitable characteristics (i.e. human data, recruitment at time of challenge/infection, regular (daily) swabbing, pathogen on the multiplex array, publicly available individual trajectory data). We searched across multiple databases for terms ("viral kinetics" AND ("SARS*" OR "CoV" or "Corona*") AND "human"), with similar searches for influenza, rhinovirus and RSV. Two studies [8, 9] were identified containing data on RSV and Influenza A viral trajectories in humans, however the viral concentration was reported in the unit of TCID$_{50}$ (tissue culture infectious dose), which cannot be easily converted to the unit required when modelling LFD and PCR test sensitivity, which is viral copies ml$^{-1}$. Several other studies (7) provide graphical summaries of individual human data or average concentrations from studies or further context, but data extraction was a barrier to usage.

To model LFD test sensitivity, the central estimate of the fitted logistic relationship between gene copies ml$^{-1}$ and LFD test sensitivity was extracted from Eyre et. al. [5]. For PCR testing, there are no estimates of the relationship between test sensitivity and gene copies ml$^{-1}$. Instead, the performance of PCR tests is often defined in terms of the test's limit of detection (LoD). If the viral concentration of the sample is below the LoD then the test is unable to confidently detect the presence of the pathogen, and a negative result will be returned for that sample. There is substantial variation in the LoD across different PCR tests, however for our model of PCR test sensitivity we have assumed an LoD of 500 gene copies ml$^{-1}$, though this was varied between 100 and 1000 as part of a sensitivity analysis. If a sample contains a viral concentration above the LoD, then the PCR test is assumed to have a test sensitivity of 95%, which was varied between 90% and 99% as part of a sensitivity analysis.

To simulate outbreaks, estimates of the generation time distribution and the incubation period for SARS-CoV-2 and influenza A/B were obtained from the literature. The incubation and generation time parameters are provided in Table (1).

| Parameter | Pathogen | Distribution | Mean (days) | Variance (days) | Source |
|---|---|---|---|---|---|
| Generation time distribution | SARS-CoV-2 | Gamma, shape = 1.81, rate = 0.455 | 3.98 | 8.74 | [10] |
| | influenza A | Gamma, shape = 3.77, rate = 1.41 | 2.67 | 1.9 | [11] |

| | | | | | | |
|---|---|---|---|---|---|---|
| | influenza B | Assumed to be the same as Flu A due to a lack of data | | | | |
| Incubation period distribution | SARS-CoV-2 | Gamma, shape = 5.81, rate = 1.05 | | 5.53 | 5.26 | [10] |
| | influenza A | Lognormal, logmean = 0.336, log standard deviation = 0.412 | | 1.71 | 1.49 | [12] |
| | influenza B | Assumed to be the same as Flu A due to lack of data | | | | |

*Table 1: Summary infectious disease parameters*

### Viral concentration trajectories

It is important to note the difference between viral load, which refers to a measure of the amount of virus present in an infected individual, and viral concentration, which is a measure of the concentration of viral particles present in a sample. As a result, viral concentration is a measure of the concentration of virus present in a swab of an individual's nasal or oral cavity and acts as a proxy measure to the viral load of a case, as it depends on both the quality of the swabbing technique and the efficiency of the release of the virus from the swab into the transport medium. Other reports and papers may refer to viral concentration as viral load. Viral concentration is the quantity of interest for modelling test sensitivity, as it specifically refers to swabs. Here, viral concentration trajectories are used to model test sensitivity over time for SARS-CoV-2 and influenza A/B cases.

For the viral concentration trajectory model, a piecewise linear model on the $\log_{10}$ gene copies ml$^{-1}$ scale with two components was used. The first component describes the initial rise in viral concentration following infection, and the second component describes the decay in viral concentration after the viral concentration peak has been reached. Given that this model is on the $\log_{10}$ scale, it is equivalent to assuming exponential growth in the viral concentration, followed by exponential decay in the viral concentration.

Each viral concentration trajectory is defined by the following triplet: peak viral concentration $p \in \mathbb{R}_+$, delay from infection to reaching peak viral concentration, $d_{i2p} \in \mathbb{R}_+$, and delay from reaching peak viral concentration to achieving viral clearance $d_{p2c} \in \mathbb{R}_+$. For a dataset with $N_{\text{cases}}$ cases, let $\underline{p}, \underline{d}_{i2p}, \underline{d}_{p2c}$ be the vectors of viral concentration trajectory parameters associated with each case. These parameters are modelled using:

$$\underline{p} \sim \text{lognormal}(\mu_p, \sigma_p^2),$$

$$\underline{d}_{i2p} \sim \text{gamma}(\alpha_{i2p}, \beta_{i2p}),$$

$$\underline{d}_{p2c} \sim \text{gamma}(\alpha_{p2c}, \beta_{p2c}),$$

where $\mu_p, \sigma_p^2, \alpha_{i2p}, \beta_{i2p}, \alpha_{p2c}, \beta_{p2c}$ are parameters to be inferred. For the SARS-CoV-2 viral concentration trajectories, the following weakly informative prior distributions are used;

$$\mu_p \sim \mathcal{N}(log(9), 3),$$

$$\sigma_p \sim \mathcal{N}(0,3),$$

$$\alpha_{i2p} \sim \mathcal{N}(4,2),$$

$$\beta_{i2p} \sim \mathcal{N}(4,2),$$

$$\alpha_{p2c} \sim \mathcal{N}(6,2),$$

$$\beta_{p2c} \sim \mathcal{N}(4,2).$$

For the influenza A/B viral concentration trajectories, the following weakly informative priors were used;

$$\mu_p \sim \mathcal{N}(log(5),1),$$

$$\sigma_p \sim \mathcal{N}(1,1),$$

$$\alpha_{i2p} \sim \mathcal{N}(5,3),$$

$$\beta_{i2p} \sim \mathcal{N}(4,1),$$

$$\alpha_{p2c} \sim \mathcal{N}(7,2),$$

$$\beta_{p2c} \sim \mathcal{N}(5,1).$$

Let $\tau \in \mathbb{R}$ be the infectious age of the case, defined as the amount of time that has passed since their time of infection. The viral concentration of the case, denoted by $v$, in $\log_{10}$ gene copies ml$^{-1}$ is given by the following function:

$$v = f(\tau, p, d_{i2p}, d_{p2c}) = \begin{cases} \tau \dfrac{p}{d_{i2p}}, & \tau \leq d_{i2p} \\ p - (\tau - d_{i2p}) \dfrac{p}{d_{p2c}}, & \tau > d_{i2p} \end{cases}$$

In the Human Challenge Study [6], cases were deliberately inoculated with virus, and consequently their time of infection is known. Therefore, in the human challenge study data for each observation of a cases viral concentration the infectious age $\tau$ is known, and given viral concentration trajectory parameters $p, d_{i2p}, d_{p2c}$, the likelihood for an observation of the viral concentration $y$ can be calculated as

$$y \sim \mathcal{N}(f(\tau, p, d_{i2p}, d_{p2c}), \sigma),$$

where $\sigma \in \mathbb{R}_+$ is an error term to be estimated. Negative PCR test results are treated as censored observations of the viral concentration trajectory, as it implies the viral concentration in the sample was below the limit of detection of the PCR test, assumed to be a Ct value of 40. As such, the likelihood of observing a negative observation of the viral concentration is given by $\mathbb{P}(y > 40)$.

For the study performed by Lau et. al. [7], cases were recruited from a household study after an index case in their household tested positive. Consequently, the time of infection is an unknown, unlike in the Human Challenge study, and as such it is necessary to estimate the time of infection as part of the model. To achieve this, analysis is restricted to cases that had a symptom onset date, and each observation of the viral concentration is now associated with a 'days from symptom onset' value, rather than a 'days since infection' value. The incubation period is estimated for each case, which provides the infectious age of the case at time of symptom onset, i.e. days_since_infection = incubation_period + days_from_symptom_onset. This allows us to use the previously discussed viral concentration trajectory model, which is parameterised in terms of days since infection.

## Modelling outbreaks and testing interventions

This analysis is focussed on modelling testing policies for the first five symptomatic cases in an outbreak in a setting up until the time that the first case is detected. As such, it is not necessary for us to simulate entire epidemic trajectories. A branching process model is therefore appropriate to use when simulating small outbreaks. Branching processes are frequently used in infectious disease modelling as they approximate the initial phase of an outbreak when there is little depletion of susceptible individuals. As this analysis is only concerned with detection of outbreaks of small numbers of cases, it is likely that this approximation holds well for most settings. It will also likely hold true as an approximation if decision to test does not start with the first symptomatic case, and instead occurs after, e.g. the second or third symptomatic case.

It is assumed that outbreaks are seeded by a single initial case, and that the number of secondary infections from a case is independently and identically distributed from a Poisson($R_0$) random variable, where $R_0$ is the assumed expected number of secondary infections from each infection. For the main analyses, results are presented using an assumed value of $R_0 = 1.5$, however this was varied as part of a sensitivity analyses.

If a simulated outbreak only contained asymptomatic cases, then the testing strategy is never initiated, and as such, fully asymptomatic outbreaks are not relevant to the analysis of different testing strategies. Further, a detected outbreak is often defined as a laboratory confirmed case and one epidemiologically linked suspected case. The presence of a second epidemiologically linked case is used as an indicator that onwards transmission has occurred. However, it is likely that some infection control interventions will be enacted upon the detection of the first case, without waiting for a second case to be discovered. As such, this modelling focusses upon the time until the first case is detected.

The generation time distribution, which describes the distribution of delays from a case becoming infected to transmitting the pathogen, is used to sample the times at which an infection creates new cases in the branching process. This allows us to appropriately describe the speed of the epidemic, which is important when attempting outbreak control, however, this results in a model where the time at which a case transmits infection is independent of their viral concentration trajectory. Cases are more likely to transmit the virus when their viral concentration is high, however there are no datasets that would allow us to estimate the relationship between viral concentration and transmission. As such, it is necessary to simulate the transmission times according to the generation time distribution.

It is assumed that the probability of a case being asymptomatic is 0.33 across all pathogens, due to a lack of data regarding the frequency of asymptomatic influenza infections - particularly in key populations of interest such as care home residents. This parameter has been varied as part of a sensitivity analysis. Symptomatic cases have a symptom onset delay drawn from an incubation period distribution for each infection.

For this initial analysis the following testing strategies are modelled: all LFD, all PCR, all concurrent, confirmatory PCR, and all LFD with PCR retesting. Brief descriptions of these strategies are provided in Table (2).

Throughout this analysis, it was assumed that LFD tests and PCR sample collection kits would be immediately available. This assumption would be met if outbreak settings have a testing/sample collection kit stockpile, however it may be the case that this is not achievable and tests need to be delivered to outbreak settings, which would necessarily incur some delivery delay.

| Strategy | Description |
|---|---|
| All LFD | - The first five symptomatic cases are tested using LFD tests<br>- Time of detection is when the first positive LFD test result is observed |
| All PCR | - The first five symptomatic cases are tested using PCR tests<br>- The time of detection is when the first positive PCR test result is observed |
| Concurrent testing | - The first five symptomatic cases are first tested using LFD tests<br>- If a symptomatic case returns a negative LFD test, then they are immediately retested using a PCR test<br>- Time of detection is the first time that either a positive LFD or positive PCR test result is observed |
| LFD with confirmatory PCR | - The first five symptomatic cases are tested using LFD tests<br>- When a case tests positive on LFD, confirmatory PCR testing is performed where the case is immediately retested using a PCR test<br>- For this strategy, we report both time of first positive LFD, and time of confirmed PCR positive |
| All LFD with PCR retesting if all negative | - The first five symptomatic cases are tested using LFD tests<br>- If all five LFD tests are negative, then the first five symptomatic cases are retested using PCR tests<br>- The time of detection is the first time that either a positive LFD or a positive PCR test is observed |

*Table 2: Brief summary of the different modelled testing strategies*

The all-LFD strategy tests the first five symptomatic individuals in a suspected outbreak using a multiplex LFD test. The time of detection for the outbreak is then the time of the first positive test result.

The all-PCR strategy, which reflected the strategy in care homes at the time of performing this analysis, is very similar to the all-LFD testing strategy; the first five symptomatic individuals in an outbreak are tested with multiplex PCR tests. However, a key feature of PCR tests is that they need to be collected and processed at laboratory. Therefore, it is assumed that the test kits are available on-site, and that there is a two-day delay during which the test is collected from the outbreak setting and delivered to a laboratory to be processed. Once the test has been processed the test result is immediately available and, if positive, used to declare an outbreak. If a new individual develops symptoms while another individual is awaiting the results of their test, then they will also be tested and these tests contribute to the total number of tests used before the outbreak is detected.

The all-concurrent strategy is a hybrid strategy, utilising a mixture of LFD and PCR tests. Symptomatic individuals are first tested with multiplex LFD tests. If the LFD test is negative, then they are retested using a multiplex PCR test. The time of detection is therefore the first time that a case either tests positive on LFD, or the first time a positive PCR test result is received. By first testing cases using an LFD test, it is possible that many outbreaks will be detected using the faster and cheaper LFD. This can allow for more rapid detection of many outbreaks, which can be important for initiating time-sensitive interventions. However, the LFD tests are known to have worse sensitivity than the PCR test, and several cases may have a false negative on the LFD test – particularly if the case has a small viral concentration at the time of testing. The aim of retesting individuals who have tested negative on LFD using PCR test is to detect these individuals with low viral concentration, and to therefore detect outbreaks that may have otherwise been missed by LFD testing alone.

The confirmatory PCR strategy is another hybrid strategy. In this strategy, symptomatic individuals are first tested using multiplex LFD tests. Individuals who test positive on LFD test are then retested using PCR tests to "confirm" their infection. Such a strategy may be desirable when the false positive rate is believed to be high, possibly due to a pathogen having very low prevalence. However, the retesting of individuals that have tested positive creates some ambiguity regarding the definition of time of detection, as it is likely that some form of action or intervention will need to take place upon the LFD result, rather than waiting for the confirmatory PCR result. For this strategy, it is necessary to distinguish between the time of first positive (which is assumed to be time of detection), and the time of confirmatory PCR result. The time of first positive is equivalent to the time of detection in the "all LFD" strategy. While a case is awaiting their confirmatory PCR test result, it is assumed that LFD testing of symptomatic cases continues, and that other positive LFD test results will also undergo a confirmatory PCR test result. This can result in several more cases undergoing PCR testing while the first LFD-positive case is awaiting the results of their confirmatory PCR test.

The final strategy considered strategy is "all LFD with PCR retesting if all negative". In this strategy, the first five symptomatic cases in an outbreak will be tested using multiplex LFD's. If all five LFD test results are negative, then the first five symptomatic cases will all be retested using PCR tests. As a result, it is possible that some of these cases will be retested with PCR some time after they were infected, at which point they have cleared the infection. Time of detection is either the first time that a case test positive on LFD, or if no symptomatic cases test positive on LFD, the time of receiving a positive PCR result from the PCR retesting. In this testing strategy, the PCR retesting primarily acts as a last attempt to detect the outbreak if all the LFD tests have returned false negative results due to poor sensitivity. While not explicitly modelled here, there are some additional benefits to this strategy. Firstly, multiplex PCR tests often have a wider panel than multiplex LFD tests, which is useful in the event that the outbreak is caused by an infectious agent not included in the multiplex LFD test panel. Without the use of PCR tests, this would result in a cryptic outbreak where the pathogen is unknown.

To compute estimates of the quantities of interest, such as time of detection, for each pathogen and strategy 1000 posterior draws of the parameters of viral concentration trajectories are used. As there is no easily obtainable analytical expression for the quantities of interest, a monte carlo estimation procedure is used. For each posterior draw, 250 outbreaks were simulated and then used to compute sample estimates of the quantities of interest, e.g. for a given posterior draw of parameters, the outbreak detection probability is computed as the fraction of 250 simulated outbreaks that were detected. The resulting estimates of the key variables of interest are consequently affected by the uncertainty in the estimated viral concentration trajectory distributions, and to a lesser extent a small amount of monte carlo error.

A sensitivity analysis was performed where the following assumed parameters were varied; asymptomatic probability, effective reproduction number, assumed model of multiplex LFD test sensitivity, and assumed PCR test sensitivity values.

# Results

## Viral concentration trajectory models

The results for fitting the viral concentration trajectory models for SARS-CoV-2 and influenza A/B are plotted in Figures (1, 2, and 3) respectively, with subpanels containing; (A) the observed viral concentration trajectory data, (B) posterior samples of viral concentration trajectories from the fitted viral concentration trajectory model, (C) the assumed model of LFD test sensitivity, (D) samples of test sensitivity over time, using posterior viral concentration trajectory samples.

SARS-CoV-2 viral concentration typically peaks at around day five of the infection, whereas influenza A/B viral concentration typically peaks at around day two of the infection. This is consistent with influenza A/B having a shorter generation time distribution than SARS-CoV-2. Additionally, SARS-CoV-2 is estimated to have a higher viral concentration peak and longer peak to clearance time than influenza A/B.

As a result of SARS-CoV-2 having a higher peak viral concentration than influenza A/B, LFD test sensitivity is higher than for influenza A/B. Influenza B appears to have the lowest viral concentration peak, which results in reduced LFD test sensitivity for some infections.

However, these models are based on relatively small numbers of observations and there is significant uncertainty regarding the parameters of these viral concentration trajectories. This is particularly the case for influenza A/B, as there are typically only three observations of the viral concentration trajectory for each case, often with a gap of several days between each observation. The fitted viral concentration trajectories for each influenza A/B case therefore have more uncertainty when compared to the SARS-CoV-2 trajectories. In addition, for influenza A/B time of infection is inferred, rather than observed as in the SARS-CoV-2 data, which creates an additional source of uncertainty. When performing simulation of testing strategies, sampling from the full model posterior is performed so that the uncertainty is fully propagated.

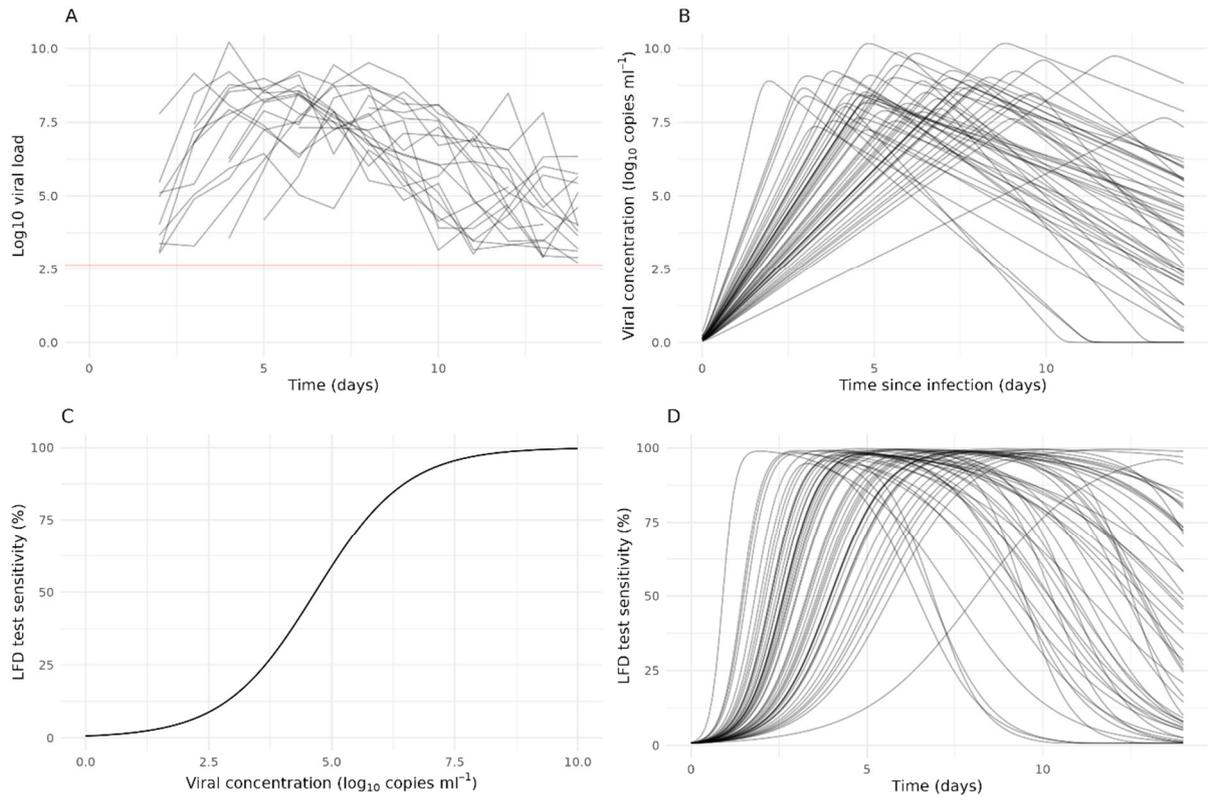

*Figure 1: SARS-CoV-2 viral concentration trajectories from fitting to the Human Challenge Study dataset for mid-turbinate swabs (A) Viral concentration trajectory data used to fit model, (B) posterior samples of viral concentration trajectories, (C) Fitted model of LFD test sensitivity, (D) posterior samples of test sensitivity over time.*

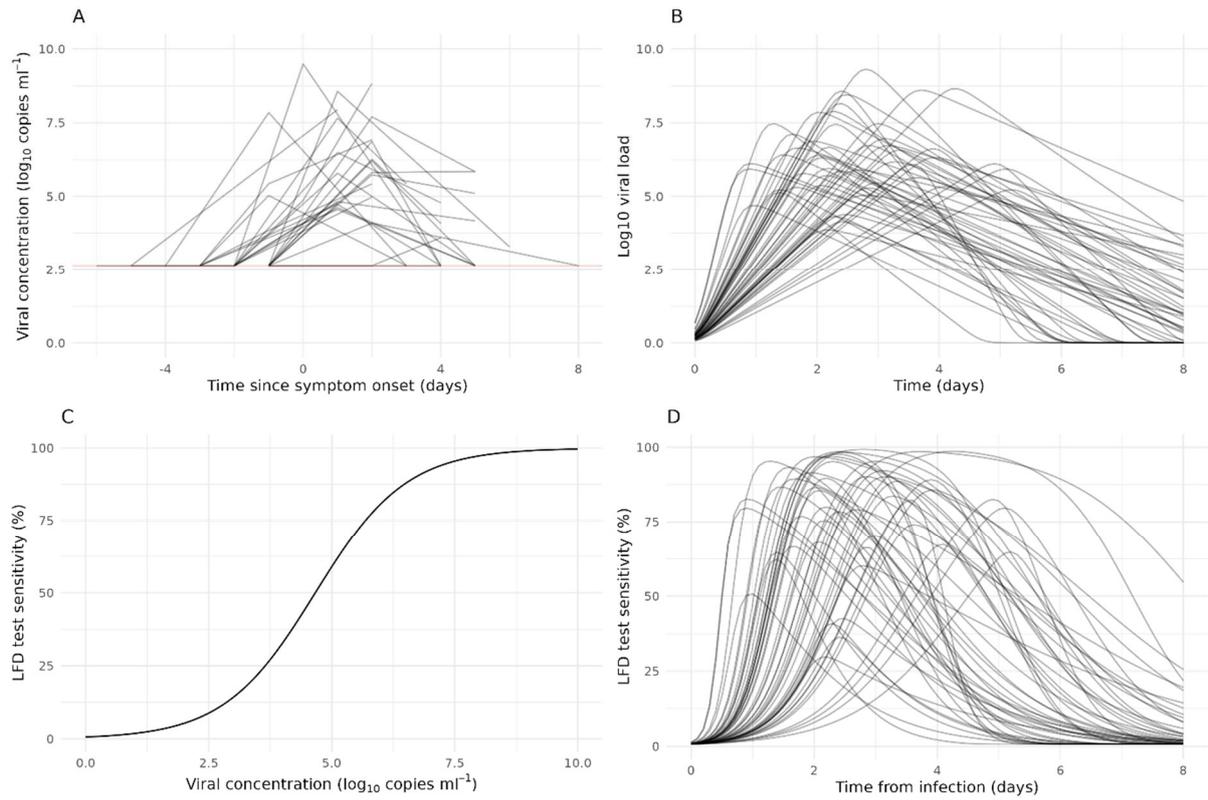

*Figure 2: Results from fitting the viral concentration trajectory models to the Influenza A cases contained in the Lau et. al. dataset. (A) Viral concentration trajectory data used to fit model, where the limit of detection is indicated by a red line. (B) Posterior samples of viral concentration trajectories. (C) Fitted model of LFD test sensitivity. (D) Posterior samples of test sensitivity over time.*

## influenza B

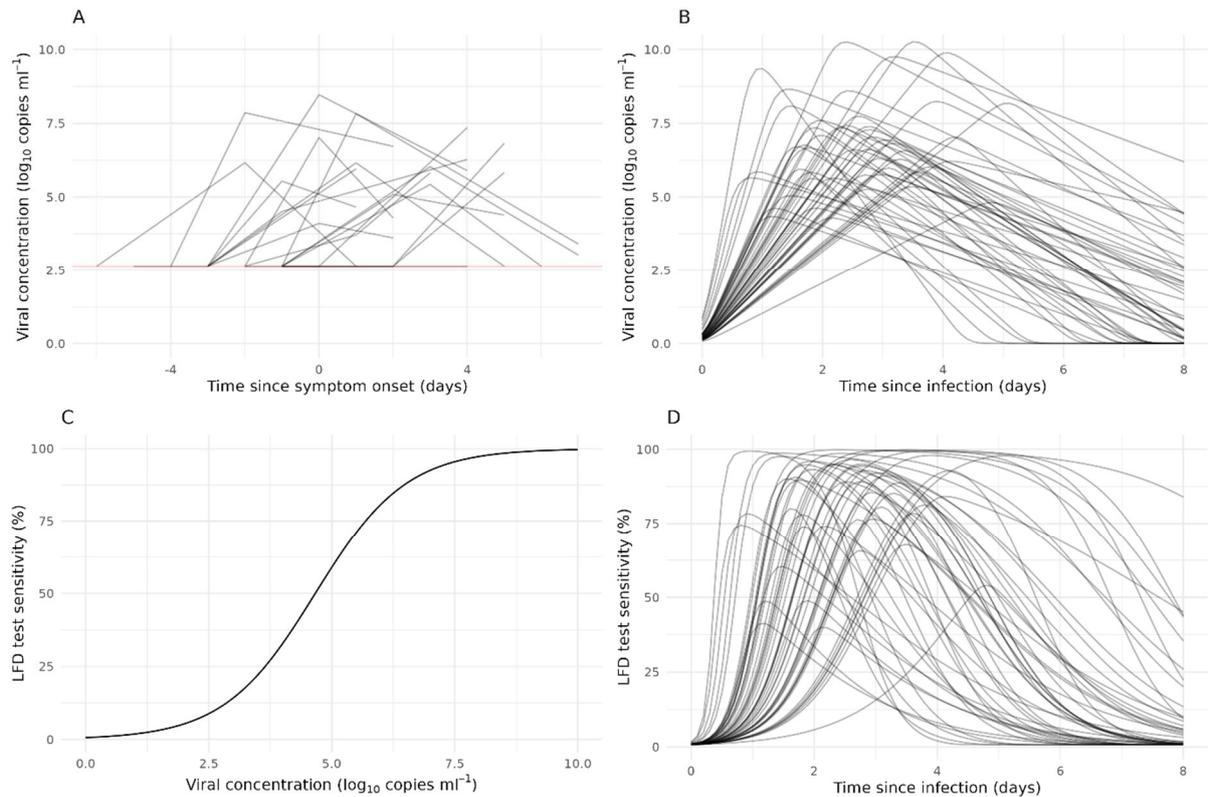

*Figure 3: Results from fitting viral concentration trajectory models to the Flu B cases contained in the Lau et. al. dataset. (A) Viral concentration trajectory data used to fit model, where the limit of detection is indicated by a red line. (B) Posterior samples of viral concentration trajectories. (C) Fitted model of LFD test sensitivity. (D) Posterior samples of test sensitivity over time*

## Results – epidemic growth curves

Before discussing the result of testing strategies, it is useful to first visualise the relative speed of each epidemic (Figure 4). This is to highlight that, while each infection is modelled using an assumed value of $R_0 = 1.5$, due to the shorter generation time distribution of influenza A/B, those viruses will spread much faster than SARS-CoV-2. One consequence of this is that delays will likely be more impactful to Influenza A/B outbreak detection strategies.

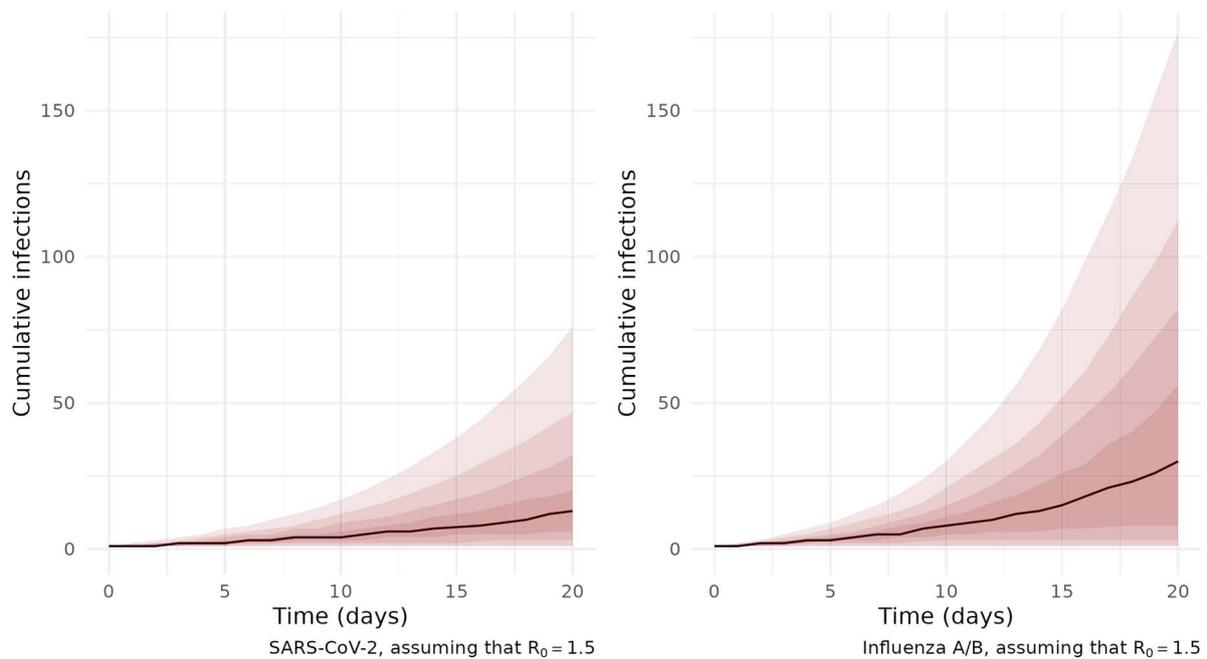

*Figure 4: Epidemic curves for SARS-CoV-2 and influenza A/B under an assumed $R_0$ of 1.5*

## Results – multiplex testing strategies

In Figures (5, 6, 7, 8, 9, and 10) comparisons of different summary statistics for each of the different testing strategies and for each infection are plotted.

### *Overview of results*

The dominant testing strategy appears to be 'concurrent testing', where symptomatic cases are first tested using LFD tests, and if a negative LFD is observed they are immediately retested using a PCR test. This provides the main benefit of LFD tests, in the form of cheap and rapidly available results for the majority of cases, while still maintaining a high outbreak detection probability due to the PCR testing of negative cases to compensate for the sometimes-poor sensitivity of LFD tests. Further, the use of LFD tests, with some PCR testing allows for the quickest detection times, and consequently the smallest numbers of total infections that have occurred by time of detection. In addition, the concurrent testing strategy uses the least amount of LFD and PCR tests, when compared to the 'all LFD' and 'all PCR' testing strategies.

*Outbreak detection probability*

The probability an outbreak is detected is given in Figure (5), where it can be observed that strategies that primarily rely on LFD tests, i.e. 'all LFD', 'all LFD with confirmatory PCR', and 'all LFD with PCR retesting', have a small chance to miss some outbreaks when compared to 'all PCR' testing and 'all concurrent' which have the highest rates of detection.

Outbreak detection rates are highest for SARS-CoV-2, followed by influenza A, and then influenza B. For SARS-CoV-2, nearly all outbreaks are detected using LFD-based testing strategies, however for influenza B a portion of outbreaks are missed by LFD-based testing strategies. The uncertainty is largest for influenza B, which is a result of posterior uncertainty in the estimates of the viral concentration trajectory distribution parameters.

In the case of the 'all LFD, retest with PCR if all negative' strategy, the addition of PCR retesting may improve outbreak detection rates for influenza B outbreaks, when compared to the 'all LFD' strategy. This is a consequence of the PCR retesting being triggered more often for influenza B, as there are more outbreaks with five cases that have all test negative on LFD. The 'all LFD', 'all LFD with confirmatory PCR' and 'all LFD, retest with PCR if all negative' strategies appear to miss a fraction of outbreaks, the outbreaks that were missed may only have contained one or two symptomatic cases before the outbreak ended.

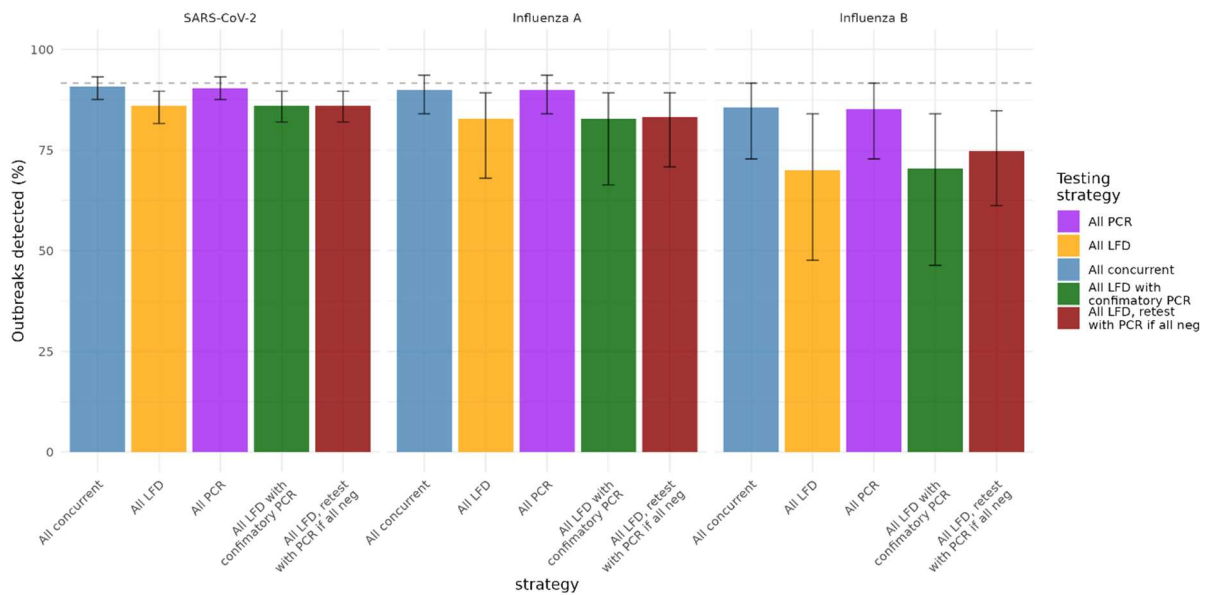

*Figure 5: Probability of detecting an outbreak (i.e. observing a single positive test, not confirmatory positive), across different testing strategies and infections. The dashed horizontal line indicates the proportion of outbreaks that died out and only contained asymptomatic cases, and therefore never underwent testing.*

In Figure (5), the proportion of detected outbreaks across all outbreaks was plotted, which included some outbreaks that either died out without a symptomatic case occurring, or outbreaks that died out after having less than five symptomatic cases. The consequences of not detecting an outbreak which ultimately contained few cases, e.g. one or two cases, are less severe than not detecting an outbreak that contained more than 5 symptomatic cases as the latter outbreak would likely have a much larger final size. Figure (6) plots the probability that an outbreak contained at least five symptomatic cases and was not detected, i.e. the first five symptomatic cases all received false negatives. These are outbreaks which may have been detected if the strategy had been to test more than the first five symptomatic cases.

In Figure (6), it can be seen that for SARS-CoV-2 and influenza A, it is very rare for outbreak with five symptomatic cases to go undetected, subject to a small amount of uncertainty in the case of influenza A. However, for influenza B, under the 'all LFD' and 'all LFD with confirmatory PCR' around 9-10% of outbreaks with at least 5 symptomatic cases would go undetected after five LFD tests and could go on to cause large outbreaks. This is subject to a significant amount of uncertainty, arising from uncertainty in the viral concentration trajectory parameters for influenza B.

In Figure (6), the results demonstrate that the 'all LFD, retest with PCR if all negative' improves on the outbreak detection rate for influenza B when compared to the 'all LFD' strategy, as there are a number of outbreaks that remain to be detected after the first five LFD tests. For influenza A and SARS-CoV-2, the additional PCR retesting testing does not appear to substantially alter the detection rate.

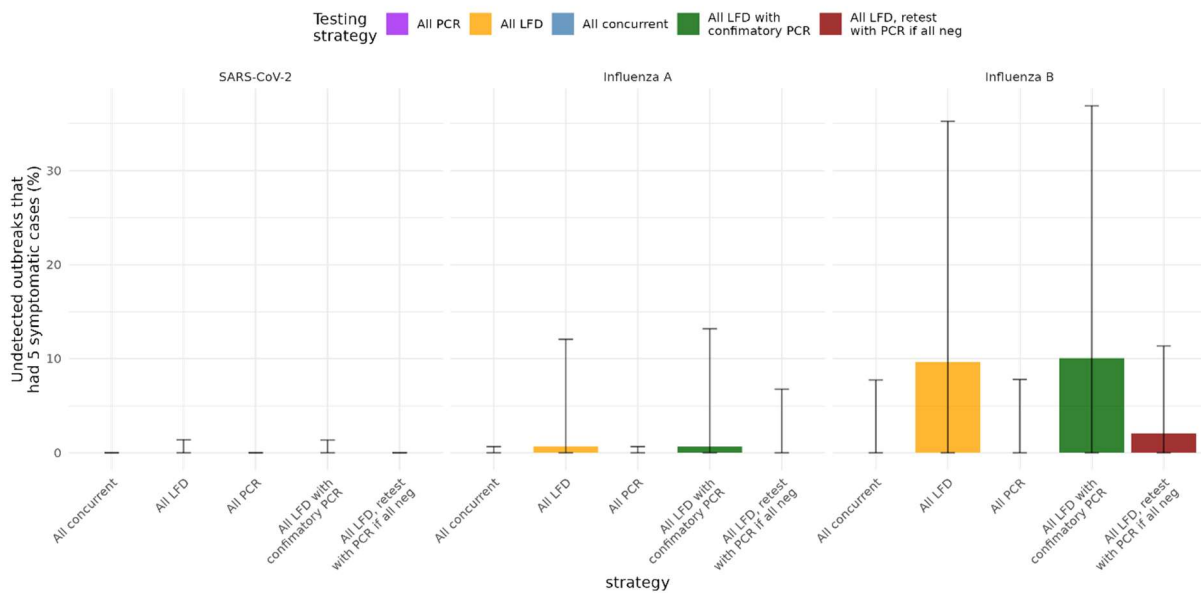

*Figure 6: Fraction of outbreaks that were not detected, given that the outbreak had 5 symptomatic cases*

### Size of outbreak before it is detected

Figures (7) and (8) visualise the time of detection and the total number of infections that have occurred by the time of detection. These are highly related plots, as a faster time of detection naturally results in fewer total infections having occurred by the time the outbreak is detected.

Figure (7) demonstrates that the time of detection was fastest for influenza A, followed by influenza B, followed by SARS-CoV-2. SARS-CoV-2 is often detected later on due to its longer generation time distribution and longer symptom onset delay. In terms of the total number of infections that had occurred by time of detection, on average SARS-CoV-2 had the most cases, followed by influenza B, followed by influenza A.

The results in Figures (7) and (8) demonstrate that concurrent testing offers the fastest time of detection, followed by 'all LFD', 'all LFD with confirmatory PCR' and 'all LFD, retest with PCR if all neg'. The slowest strategy to detect cases is 'all PCR', primarily due to the two-day delay in receiving test results.

For influenza A/B, switching from the 'all PCR' testing strategy to the 'all concurrent' testing strategy would nearly halve the total number of infections that had occurred by time of detection. While the primarily LFD based testing strategies offer fast time of detection, it is worth recalling that some outbreaks were not detected by these strategies, though most of the undetected outbreaks would contain less than five symptomatic cases.

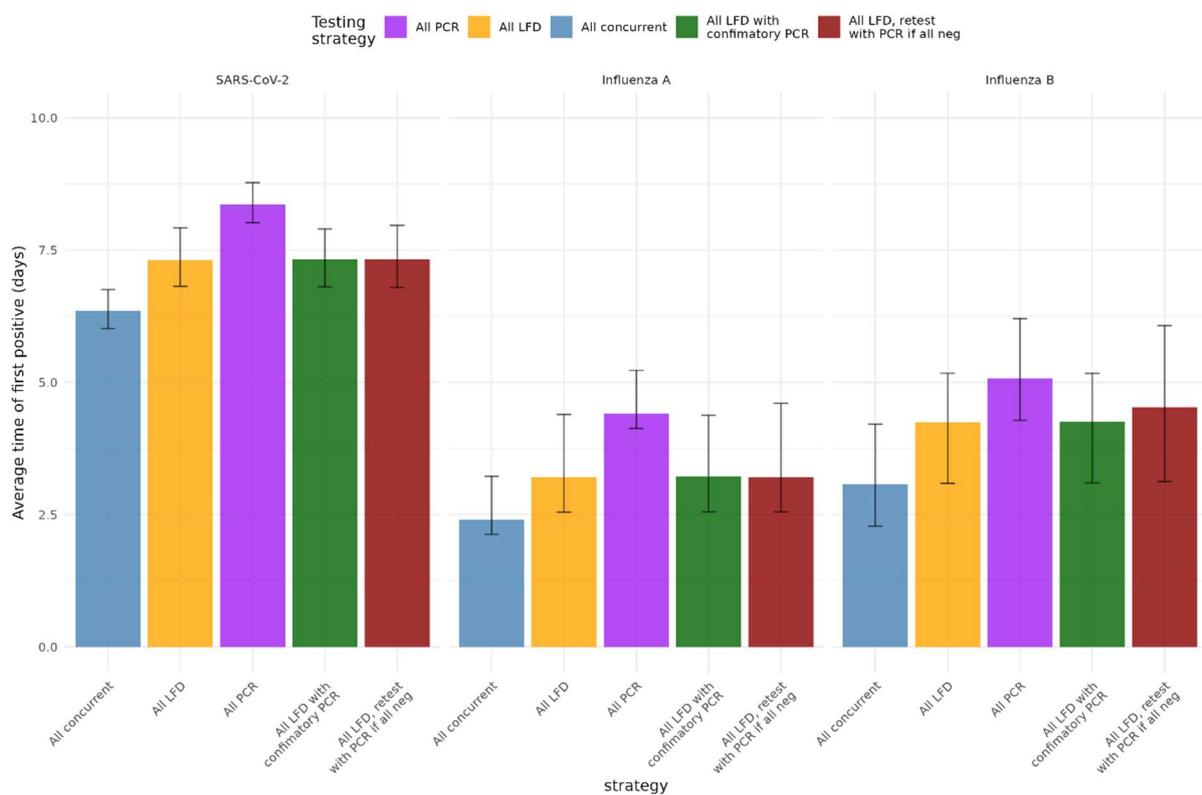

*Figure 7: The average time of observing the first positive test for different testing strategies.*

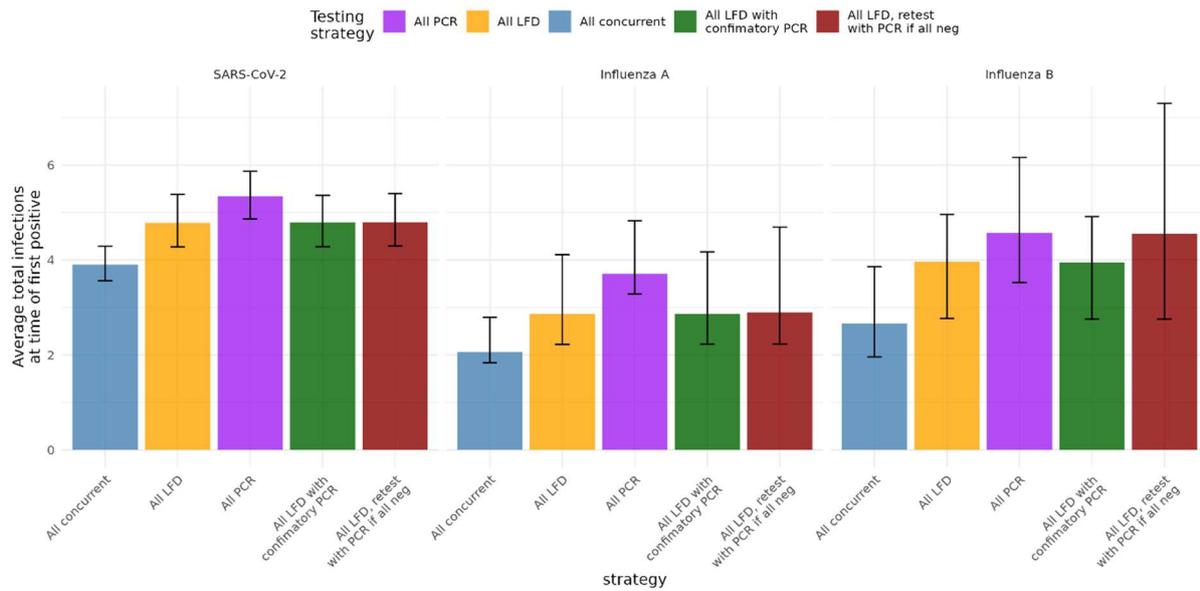

*Figure 8: The average number of infections that have occurred by the time a positive test result has been observed.*

*Confirmatory PCR testing impact*

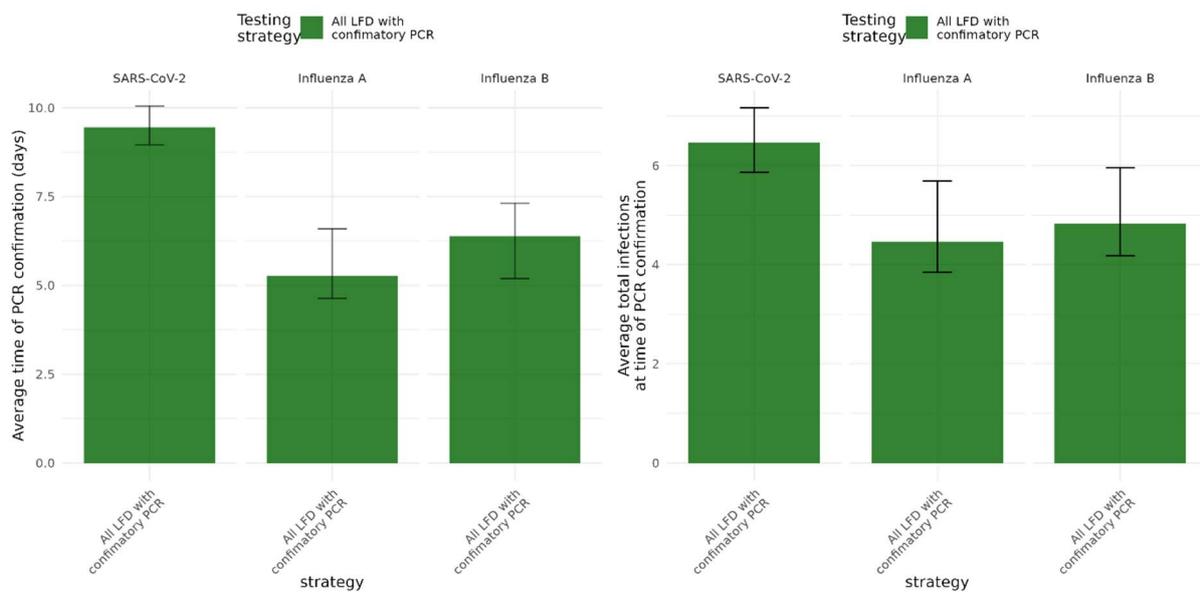

*Figure 9: (left) the time of receiving PCR confirmation of an outbreak, and (right) the number of cases at time of PCR confirmation.*

For the strategy, 'All LFD with confirmatory PCR', it is assumed that testing of the first five symptomatic cases continues until a positive confirmatory PCR result is observed. As such, it is of interest to understand how many tests were used by time of first positive, which was plotted in Figures (7 and 8), and at time of PCR confirmation, which is plotted in Figure (9). During the assumed two-day delay in receiving PCR test results, the outbreaks continue spreading, resulting in more cases at time of PCR confirmation when compared to time of first positive, especially for influenza A/B due to their faster growth rate under the assumptions of the model.

*Number of tests used to detect outbreaks*

From a logistical requirements perspective, it is necessary to explore how many tests are utilised by each testing strategy up until observing the first positive test result, as this impacts the cost of each strategy.

The number of tests used by each strategy up until the time of detection is plotted in Figure (10). Naturally, the 'all PCR' testing strategy uses the most PCR tests, as PCR testing continues while awaiting the results of PCR tests that were performed earlier. The 'concurrent testing' strategy only uses a very small number of PCR tests as the majority of cases are detected via LFD test, with only a small number of false negatives requiring PCR retesting. Similarly, the 'all LFD, retest with PCR if all negative' strategy only uses a small number of PCR tests on average, as most outbreaks are first detected on LFD tests, except for influenza B, where reduced LFD test sensitivity leads to the PCR retesting occurring.

For the number of LFD tests used, there is relatively little variation between the different strategies, with the exception of the 'concurrent testing' strategy, which appears to use slightly less LFD tests than the other strategies.

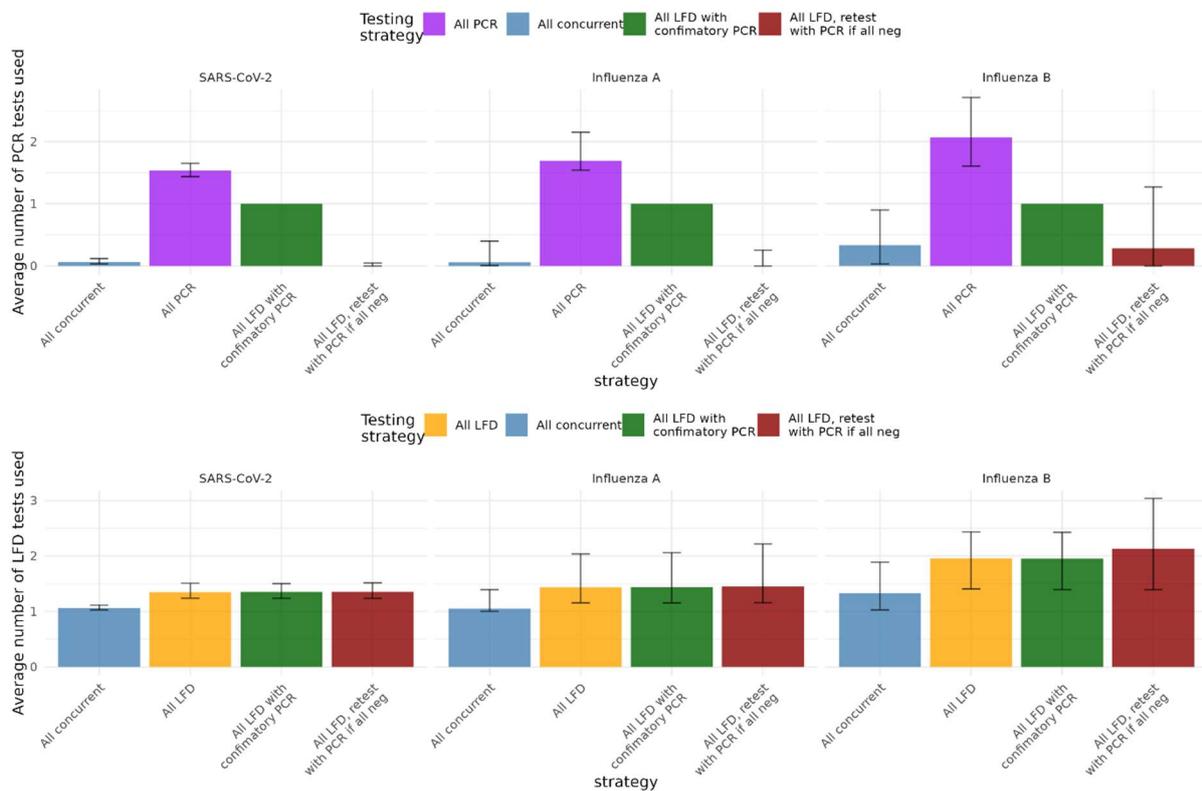

*Figure 10: (upper) the average number of PCR tests used by each strategy in detecting an outbreak, and (lower) the average number of LFD tests used by each strategy in detecting an outbreak*

Finally, Figure (11) plots the expected number of PCR tests used while awaiting confirmatory PCR testing results. For SARS-CoV-2 and influenza B, another 0.4 PCR tests will be used on average, and for influenza A another 0.5 PCR tests will be used while awaiting PCR test results.

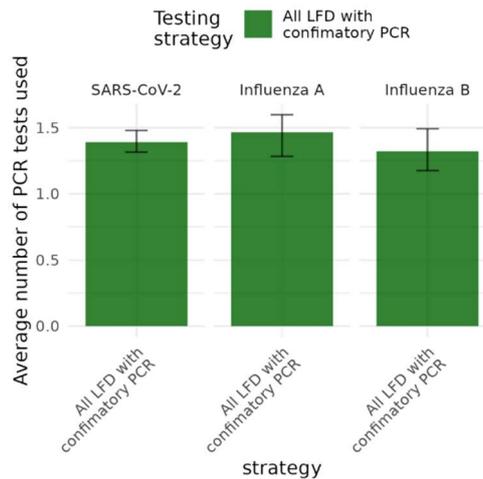

*Figure 11: The average number of PCR tests used while awaiting confirmatory PCR results.*

*Sensitivity analysis*

Several assumed parameters were varied during a sensitivity analysis; the LoD of the PCR test, the PCR test sensitivity given that the sample was over the limit of detection, the $R_0$ value, the probability that an infection is asymptomatic, and the sensitivity of the LFD test. Strategies that used PCR tests were found to be largely insensitive to changes in both the LoD of the PCR test and the sensitivity of the PCR test given that the viral concentration of a sample was over the LoD.

Decreasing the asymptomatic probability increases the amount of symptomatic testing, which increases the detection rate of outbreaks. A lower probability of asymptomatic infections also reduces the number of infections that occur by time of detection or PCR confirmation. Increasing $R_0$ also increases the amount of symptomatic testing as outbreaks tend to be larger and increases the number of infections that occur before the outbreak is detected. Changes to these epidemiological parameters generally did not change the performance of the modelled strategies relative to each other.

The LFD test sensitivity was changed by applying a shift on the $\log_{10}$ viral copies ml$^{-1}$ scale, Figure S1. It was found that the performance of strategies that primarily use LFD tests was impacted, particularly for Influenza B which was estimated to have lower viral load peaks than SARS-CoV-2 and Influenza A. There was, however, significant uncertainty associated with this finding, largely resulting from uncertainty in the Influenza B viral load trajectory.

# Discussion

Overall, the results suggest that multiplex LFD tests can be effectively deployed to detect SARS-CoV-2 and influenza A/B outbreaks. Care should be taken to ensure that multiplex LFD tests have a sufficient level of Influenza B test sensitivity however. According to the modelled metrics, it appears that the best strategy is the concurrent testing strategy, which uses each test according to the strengths of that test; initial use of rapid tests to rapidly detect high viral concentration cases, and PCR testing to detect cases with small viral concentrations. Solely relying on LFD tests may lead to some outbreaks not being detected, whereas solely relying on PCR tests can add unnecessary delays into the process and diminish the effectiveness of subsequent interventions.

In terms of testing resource, the concurrent testing strategy appears to use relatively few tests in comparison to other strategies when detecting a SARS-CoV-2 or influenza A/B outbreak. Due to the cost of PCR tests, this may result in a more expensive strategy than some of the other strategies that were primarily LFD based, though such strategies could miss a small number of outbreaks. The number of tests used by each strategy depends on the assumed effective reproduction number and test sensitivity model, which were varied as part of a sensitivity analysis. Additionally, the overall number of tests consumed by a given strategy will depend on the prevalence of similar respiratory infections at that point in time, for example, a high prevalence of Rhinovirus in the community, which may not be included in the multiplex test panel, would potentially cause lots of testing to be initiated without leading to any outbreak control interventions occurring. However, it is worth noting that multiplex LFD's and PCR's will likely reduce overall testing demand and burden by testing for multiple pathogens in a single test, rather than potentially requiring several distinct singleplex tests to work out which pathogen is causing the outbreak.

There are several important caveats to this modelling study. Due to a lack of trial data on multiplex LFD test sensitivity at the time of performing this study, the model of test sensitivity given viral concentrations for influenza A/B was assumed to be the same as for SARS-CoV-2, using the Innova SARS-CoV-2 LFD test which was the best performing test at the time of performing this analysis. The actual performance of a multiplex LFD testing strategy depends strongly on the performance of the multiplex LFD test, and SARS-CoV-2 LFD sensitivity achieved by manufacturers has been highly variable [5], suggesting that there may be a similarly variable range of performance characteristics for multiplex LFD tests. It may also be important to consider the consequences of a multiplex test where the test sensitivity to different pathogens requires a trade-off, for example a multiplex test with good influenza sensitivity and weaker SARS-CoV-2 test sensitivity.

The negative impacts of false positives have not been considered as the prevalence of influenza-like illness that would initiate testing, but which would not be included in a multiplex testing panel, is unclear. This is an important concern, given that under the concurrent testing strategy a symptomatic case that is not infected with SARS-CoV-2 or influenza A/B would be tested twice, which could increase the false positive rate. However, it is important to note that the false positive rate of PCR and LFD tests is very low, at approximately 1 in 1000.

The viral concentration trajectory models are fitted to relatively small amounts of data, particularly for influenza A/B where there are few observations per case. This leads to a significant amount of uncertainty in the posterior estimates of the parameters of the viral concentration trajectories, which has an impact on the efficacy of the modelled strategies. In general, there is relatively little data available regarding the viral concentration trajectories, and further research is required to validate the estimates of test sensitivity profile over the course of a SARS-CoV-2 or influenza A/B infection. For other pathogens of interest, such as RSV, there was no viral concentration trajectory data publicly available to fit the model to.

The assumed model of outbreaks does not allow for the possibility of overlapping outbreaks. This may be particularly problematic when there is an outbreak of SARS-CoV-2/influenza A/B, and a different infection with similar symptoms co-occurring at the same time. In addition, the model of SARS-CoV-2 best represents wild-type SARS-CoV-2, due to the variant used in the Human Challenge Study. There are likely some differences between the wild-type viral concentration trajectories and those of the currently dominant variant. Additionally, the immunity landscape has continued to evolve as new vaccines are deployed, and individuals are exposed to a range of different variants. High quality viral concentration trajectory data is not routinely collected however, so it is difficult to provide more recent viral concentration trajectories. Similar caveats apply to the model of influenza A/B, which are continuously evolving pathogens, however the dataset that was used is from 2008 as no more recent data is available.

The analysis ends once a case is detected and does not explore the actions or interventions applied once an outbreak has been detected. Therefore, it is outside the scope of this analysis to determine the costs associated with not detecting an outbreak. However, the analysis does not suggest any reasons that detecting the outbreak via LFD or PCR would affect the actions that are taken after.

The assumption that multiplex LFD/PCR tests were immediately available implies a stockpile of testing resources available at each outbreak setting. In practice, there may be logistical concerns meaning that this is not feasible, and it would be necessary to explore the impacts of delays in delivering tests to symptomatic cases.

Overall, the results show that multiplex LFD testing ought to be effective at detecting the majority of outbreaks, and the concurrent testing strategy demonstrates the potential for having strong performance. There are, however, numerous real-world consideration that need to be factored in when determining which strategy is the optimal in terms of achieving infection control goals.

# Ethical statement

UKHSA have an exemption under regulation 3 of section 251 of the National Health Service Act (2006) to allow identifiable patient information to be processed to diagnose, control, prevent, or recognise trends in, communicable diseases and other risks to public health.

# Conflicts of interest

The authors have declared that no competing interests exist.

# Acknowledgements

This work was supported by the UK government Department of Health and Social Care (DHSC), The UK Health Security Agency (UKHSA) and NHS Test & Trace. The views expressed are those of the author(s) and not necessarily those of the DHSC, UKHSA or NHS Test & Trace. We would like to thank Galena Kuyumdzhieva and Steven Dyke for assisting with literature review of viral load data.

# Data availability statement

The SARS-CoV-2 viral load trajectory data may be accessed by contacting Killingley et. al. The influenza A/B viral load trajectory may be accessed at https://github.com/bcowling/pediatric-vaccine-trial/tree/master/data.

# Code availability statement

The model code can be made available on request to DataAccess@ukhsa.gov.uk.

# Sensitivity Analysis

Several assumed parameters were varied during a sensitivity analysis; the limit of detection (LoD) of the PCR test, the PCR test sensitivity given that the sample was over the limit of detection, the $R_0$ value, the probability that an infection is asymptomatic, and the sensitivity of the LFD test. A description of the values used in the sensitivity analysis is provided in Table S1. A visualisation of the LFD sensitivity curve for different shift values is provided in Figure S1.

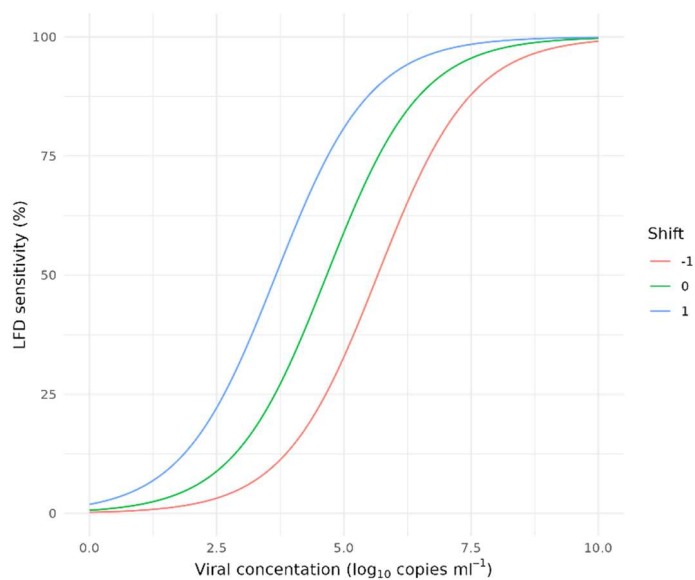

Figure S1: Visualisation of the LFD test sensitivity curve under after being shifted on the $\log_{10}$ viral copies ml$^{-1}$ scale during the sensitivity analysis

| Variable | Definition | Baseline value | Lower and upper values for the sensitivity analysis |
|---|---|---|---|
| PCR test Limit of detection | The threshold below which the PCR test will not return a positive result | $\log_{10}(500)$ viral copies per ml | $\log_{10}(100)$ and $\log_{10}(1000)$ viral copies per ml |
| PCR test sensitivity | The probability that the PCR returns a positive result, given that the sample viral concentration is above the limit of detection | 0.95 | 0.90, 0.99 |
| $R_0$ | The expected number of secondary infections from a single case | 1.50 | 1.25, 2.00 |
| Asymptomatic probability | The probability that a case is asymptomatic | 0.33 | 0.20, 0.5 |
| LFD test sensitivity shift | Shifts the models of LFD test on the log10 viral copies ml$^{-1}$ scale. This allows modelling of LFD tests that are more/less sensitive than the baseline test. | 0 | -1, 1 |

Table S1: Description of the parameters varied during the sensitivity analysis, with the baseline value that was used in the main analysis, and the two additional values that were explored during the sensitivity analysis

Figure S2 contains the quantities of interest when the probability that a case is asymptomatic is varied from it's baseline value of 0.33. As the probability of an asymptomatic case occurring increases, outbreaks which die out after a small number of infections will contain fewer symptomatic infections, resulting in less testing and a lower probability that the outbreak will be detected. Additionally, an increase in the asymptomatic probability increases the time until a symptomatic case occurs, which increases both the time to detection, and the number of infection that have occurred at time of detection, Subfigures S2C, S2D, S2G and S2H.

In Figure S3, the $R_0$ value is varied from it's baseline value of 1.5. As $R_0$ increases, the expected final size of the outbreaks increases. This implies there will be fewer outbreaks that contain less than five symptomatic cases, which increases the amount of symptomatic testing that can be performed, Subfigure S3E and S3F. An increase in the amount of symptomatic testing leads to an increased probability of detecting an outbreak, Subfigure S3A. The growth rate of the epidemic increases as $R_0$ increases, and as a result, the number of infections that has occurred at by time of detection, Subfigure S3C, and particularly for the time of PCR confimation, Subfigure S3G.

The LoD for PCR tests was varied from it's baseline value of 500 viral copies ml$^{-1}$ in Figure S4. Varying the limit of detection appears to have had a small effect on the performance of the "all concurrent" and the "all PCR" strategies, however the confidence intervals often overalap and the effect is relatively small. This suggests that the modelled outbreak detection strategies are insensitive to changes in the LoD for the PCR test, likely as a result of testing being initiated by symptom onset when the viral concentration is expected to be high. A similar narrative is observed in Figure S5, where the PCR sensitivity given that a sample was over the LoD was varied from it's baseline value of 0.95, as it appears that none of the strategies that use PCR tests are highly sensitive to changes in the PCR sensitivity value.

Finally, the results when a shift is applied to the LFD test sensitivity curve are plotted in Figure S6. The concurrent testing strategy was largely insensitive to changes in the LFD test sensitivity, likely as a result of it falling back on the PCR test. The remaning strategies, which don't feature heavy use of the PCR test, all demonstrate that the oubtreak detection probability is sensitive to a decreasing LFD test sensitivity curve, particularly for Influenza B, Subfigures S6A and S6B. The decrease in sensitivity causes small increases to the time to first positive and the number of infections at time of first positive, Subfigure S6D and S6C respectively. Naturally, a reduction in the LFD test sensitivity caused an increase in the number of LFD tests used, Subfigure S6F, and for Influenza B there was a large increase in the number of PCR tests used for the strategy "all LFD with PCR retesting if all negative", implying that the PCR retesting was triggered more frequently, Subfigure S6E.

Overall, while certain key performance indicators were sensitive to epidemiological assumptions around $R_0$ or the asymptomatic probability, and the LFD test sensitivity, the overall narrative regarding the performance of the different strategies was largely unchanged; the all concurrent testing strategy generally had the strongest performance. The performance of primarily LFD based strategies was found to be sensitive to the LFD test sensitivity curve, which is a result of Influenza B having lower viral load peaks than SARS-CoV-2 and Influenza A. Though, there is considerable uncertainty in the parameters of the Influenza B trajectory as a consequence of the small number of observations.

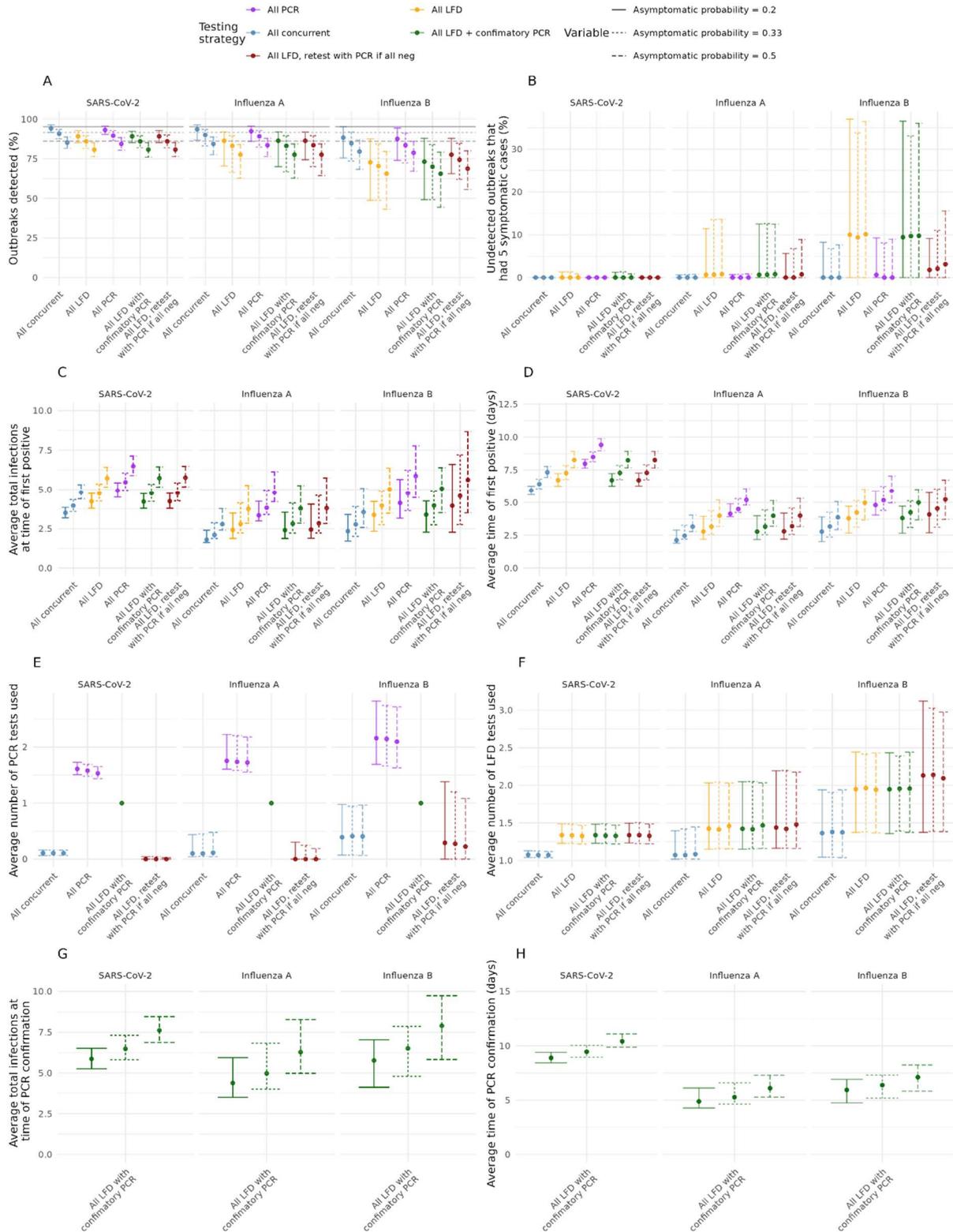

*Figure S2: Posterior estimates for the quantities of interest when the probability that an infection is asymptomatic is varied from the baseline value of 0.33. (A) The outbreak detection probability, (B) the outbreak detection probability given that there was five or more symptomatic infections, (C) the average number of infections that had occurred by time of first positive, (D) the average time of first positive, (E) the average number of PCR tests used at time of first positive, (F) the average number of LFD tests used at time of first positive, (G) the average total infections at time of PCR confirmation, (H) the average time of PCR confirmation.*

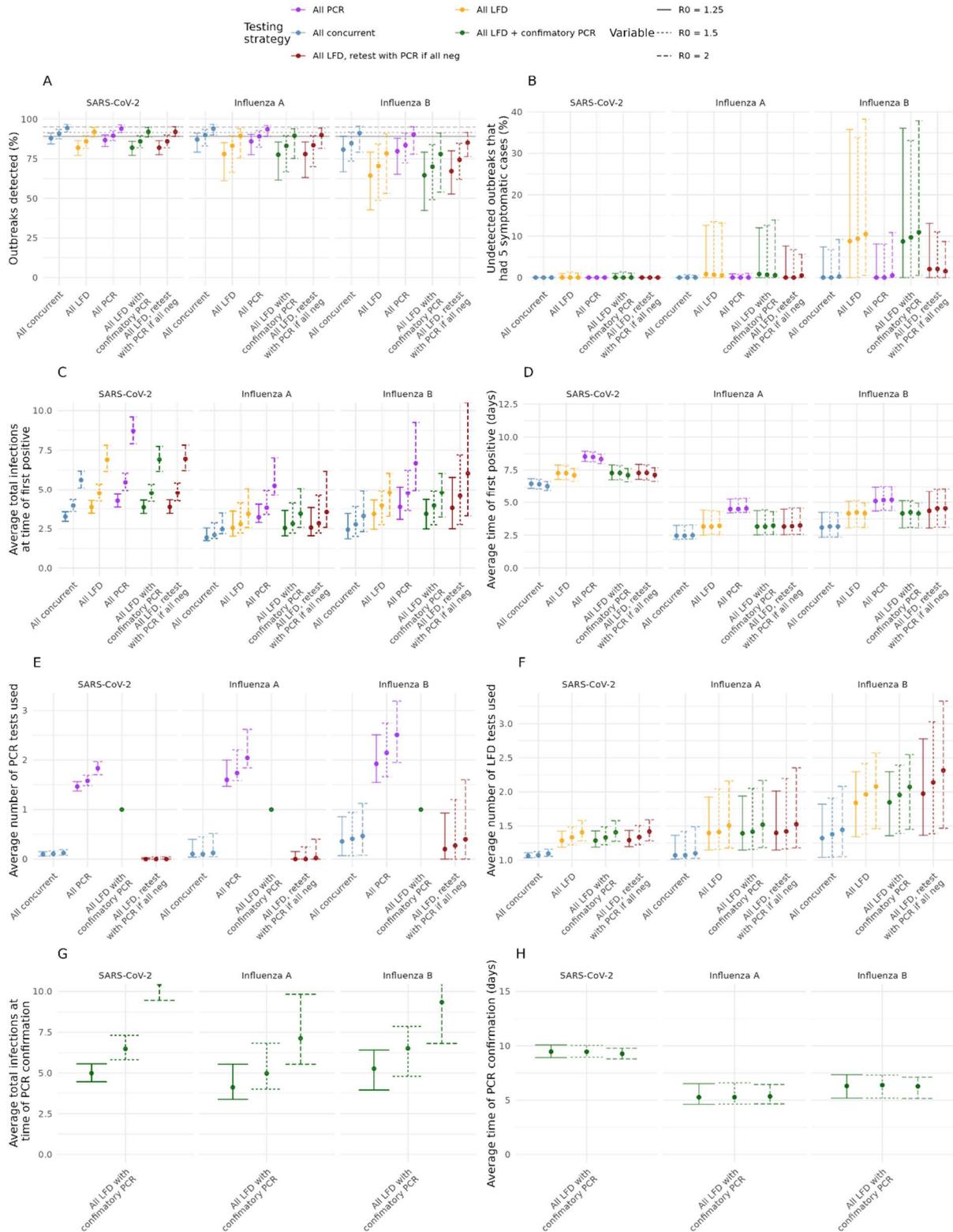

Figure S3: Posterior estimates for the quantities of interest when $R_0$ is varied from the baseline value of 1.5. (A) The outbreak detection probability, (B) the outbreak detection probability given that there was five or more symptomatic infections, (C) the average number of infections that had occurred by time of first positive, (D) the average time of first positive, (E) the average number of PCR tests used at time of first positive, (F) the average number of LFD tests used at time of first positive, (G) the average total infections at time of PCR confirmation, (H) the average time of PCR confirmation.

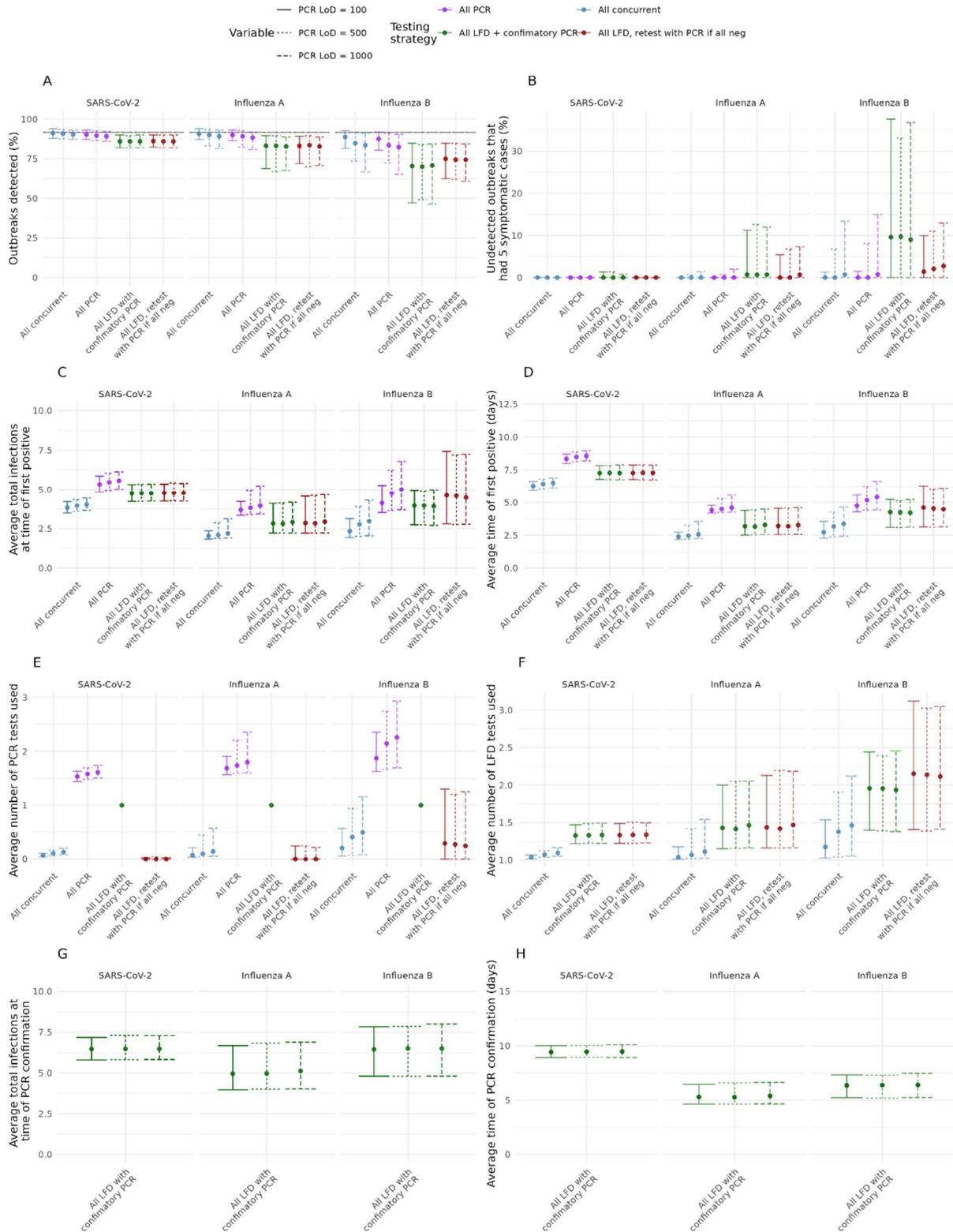

*Figure S4: Posterior estimates for the quantities of interest when the PCR LoD is varied from the baseline value of 500 viral copies ml$^{-1}$. (A) The outbreak detection probability, (B) the outbreak detection probability given that there was five or more symptomatic infections, (C) the average number of infections that had occurred by time of first positive, (D) the average time of first positive, (E) the average number of PCR tests used at time of first positive, (F) the average number of LFD tests used at time of first positive, (G) the average total infections at time of PCR confirmation, (H) the average time of PCR confirmation.*

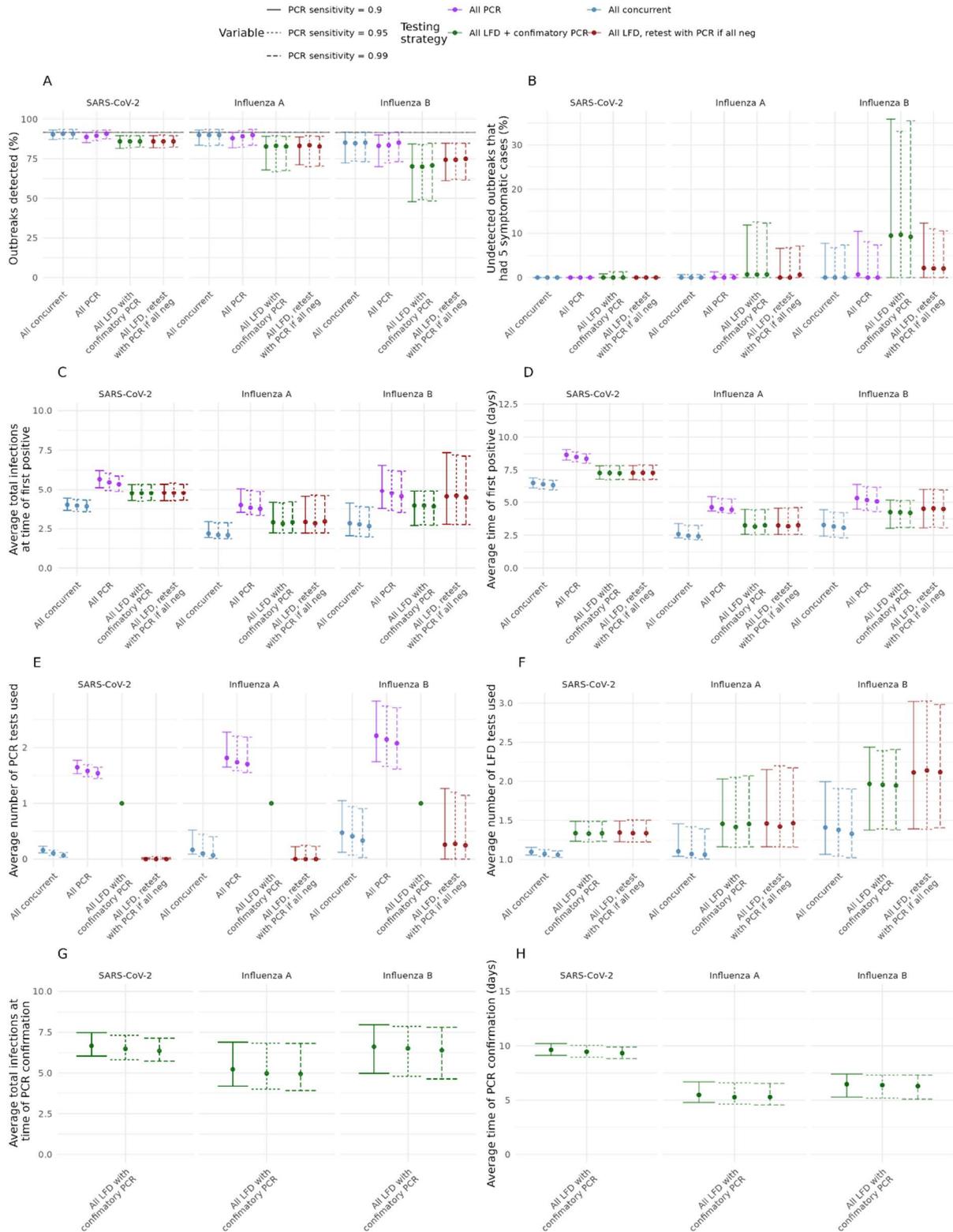

*Figure S5: Posterior estimates for the quantities of interest when the PCR test sensitivity given that a sample is above the baseline value of 95% sensitivity. (A) The outbreak detection probability, (B) the outbreak detection probability given that there was five or more symptomatic infections, (C) the average number of infections that had occurred by time of first positive, (D) the average time of first positive, (E) the average number of PCR tests used at time of first positive, (F) the average number of LFD tests used at time of first positive, (G) the average total infections at time of PCR confirmation, (H) the average time of PCR confirmation.*

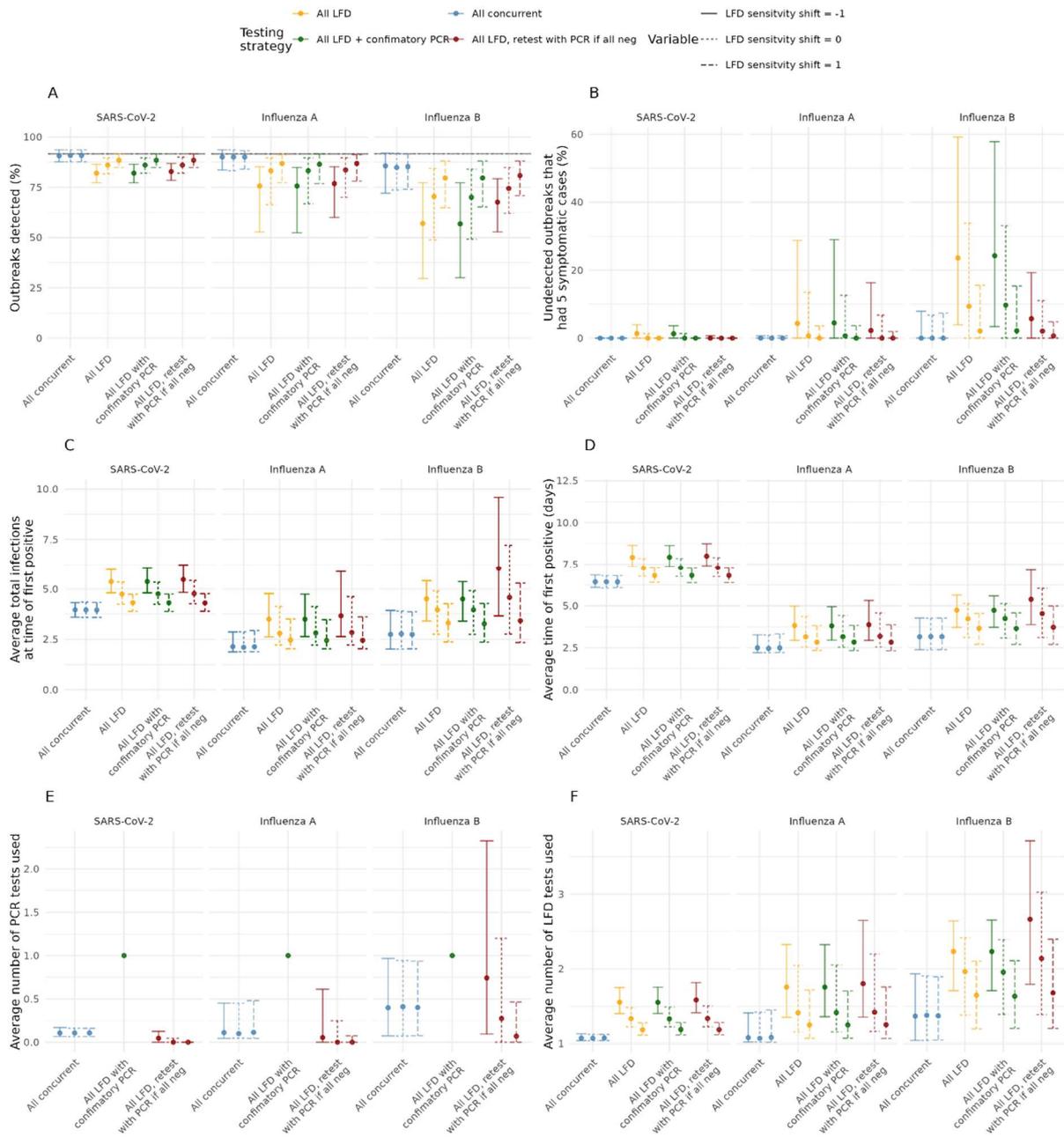

Figure S6: Posterior estimates for the quantities of interest when a shift is applied to the LFD test sensitivity on the $\log_{10}$ viral copies ml$^{-1}$ scale. (A) The outbreak detection probability, (B) the outbreak detection probability given that there was five or more symptomatic infections, (C) the average number of infections that had occurred by time of first positive, (D) the average time of first positive, (E) the average number of PCR tests used at time of first positive, (F) the average number of LFD tests used at time of first positive, (G) the average total infections at time of PCR confirmation, (H) the average time of PCR confirmation.